\documentstyle  [11pt] {article}
\title {\bf Glassy transition and metastability in four-spin Ising model \vskip5mm}   
\author {{\sc Adam Lipowski}$^{1),2)}$ and {\sc Des Johnston}$^{1)}$\\
\noalign{\vskip5mm}
$^{^{1)}}${\it Department of Mathematics, Heriot-Watt University},\\ 
{\it EH14 4AS Edinburgh, United Kingdom}\\
$^{2)}${\it Department of Physics, A.~Mickiewicz University,}\\
{\it  61-614 Pozna\'{n}, Poland}\\
\noalign{\vskip5mm}}
\date {}
\newif\ifetykiety

\def\etykieta#1{\ifetykiety \par\marginpar{\centering[#1]} \fi}
\def\bibetykieta#1{\ifetykiety \marginpar{\renewcommand{\baselinestretch}{0.9}
                   \raggedright\small[#1]} \fi}

\newcommand {\SECTION} [2] {\section{#2} \label{#1} \etykieta{#1} \setcounter 
  {equation} {0}}
\newcommand {\SUBSECTION} [2] {\subsection{#2} \label{#1} \etykieta{#1} }

\newcommand {\eq} [1] {(\ref {#1})}
\newcommand {\beq} {\begin {equation}}
\newcommand {\eeq} [1] {\label {#1} \end {equation} \etykieta{#1}}
\newcommand {\beqn} {\begin {eqnarray}}
\newcommand {\eeqn} [1] {\label {#1} \end {eqnarray} \etykieta{#1}}
\catcode `\@=11
\def\@cite#1#2{#1\if@tempswa , #2\fi}
\catcode `\@=12
\newcommand {\cyt} [1] {$^{\mbox {\footnotesize \cite{#1})}}$}

\def\bib#1#2\par{\bibitem{#1} #2 \bibetykieta{#1}}
\newcommand {\fig} [1] {Fig.~\ref{#1}}
\newcounter {fig}
\newenvironment {figure_captions} {\newpage \thispagestyle {empty} \section*
{Figure captions} \begin {list} {\bf Fig.~\arabic{fig}:} {\usecounter{fig}
\settowidth{\labelwidth}{Fig.~999:} }}{\end{list}}
\def\elem#1#2\par{\item#2\label{#1}\etykieta{#1}}

\renewcommand {\baselinestretch} {1.1}
\renewcommand {\cyt} [1] {{\mbox [\cite{#1}]}}
\begin {document}
\maketitle
\begin {abstract}
Using Monte Carlo simulations we show that the three-dimensional Ising model with four-spin
(plaquette) interactions has some characteristic glassy features.
The model dynamically generates diverging energy barriers, which give rise to slow dynamics at low
temperature.
Moreover, in a certain temperature range the model possesses a metastable (supercooled liquid) 
phase, which is presumably supported by certain entropy barriers.
Although extremely strong, metastability in our model is only a finite-size effect and sufficiently
large droplets of stable phase divert evolution of the system toward the stable phase. 
Thus, the glassy transitions in this model is a dynamic transition, preceded by a pronounced peak
in the specific heat.
\end {abstract}
\SECTION{int}{Introduction}
Recently, glassy systems have been intensively studied from both the theoretical and experimental
points of view~\cyt{GOTZE,JPHYSC}.
However, due to the very complicated nature of glasses their understanding is still far from
being complete.
Although one can construct various off-lattice models of glasses, these models are in general
very difficult to study and additional insight would be desirable~\cyt{KOB}.

One possible direction is to study lattice models of glasses in the hope that, albeit
unrealistic in some respects, these models describe at least some aspects of a glassy transition.
The advantage of using lattice models is that usually they are much easier to study.   
Indeed, one can show analytically, and in some cases even exactly, that  certain lattice models do
undergo glassy transitions~\cyt{MEANFIELD}.
However, these analytically tractable models are usually of infinite dimension or contain
interactions of infinite range and it is not certain to what extent these models are applicable to
real systems.

An interesting model exhibiting some glassy features, which is both three-dimen\-sio\-nal and
contains short-range interactions, was proposed some years ago by Shore et al.~\cyt{SS,SHORE}.
They studied the dynamics of a three-dimensional ferromagnetic Ising model with antiferromagnetic
next-nearest-neighbours interactions (the SS model).
They showed that the low-temperature coarsening of a random quench asymptotically becomes very
slow and the characteristic length scale increases logarithmically in time, which is one of
the characteristic features of glasses.
However, Shore et al. argued that their model is not yet a satisfactory model of glasses.
Namely they show that in their model the transition into the slow-dynamics regime is induced by a
corner-rounding transition. 
At this transition the energy barriers should vanish, which, according to their calculations,
implies that the zero-temperature characteristic length scale increases too fast as a
function of inverse cooling rate~\cyt{SHORE}.
Shore et al. argued that in more realistic models of glasses the energy barriers should exist
even at the glassy transition.

In our opinion, the SS model lacks yet another characteristic feature of real systems,
namely  the supercooled liquid phase.
In the SS model the high-temperature phase (liquid) cooled below the critical point $T_{{\rm
c}}$ but above the corner-rounding transition relatively quickly evolves toward the
low-temperature phase (crystal).
One would expect that when brought to this temperature range, liquid should not
immediately crystallize (or rather polycrystallize) but, 
at least for some time, it should remain in a
metastable state of supercooled liquid.

Recently, it has been shown that the three-dimensional Ising model with four-spin
(plaquette) interactions has some similarities to the SS model~\cyt{LIPO}.
Namely, it was shown that in this model certain configurations are very long-lived due to large
energy barriers.
Configurations of this kind should develop spontaneously during the coarsening and thus
should considerably slow-down this process.
Indeed, it was observed~\cyt{LIPO} that the high-temperature quench becomes trapped in a glassy phase.

In the present paper we examine this model further.
First, we show that in the low-temperature regime the characteristic length $l(t)$ increases very
slowly in time.
We argue that this increase is likely to be logarithmic, namely $l(t)\sim \ln t$.
Such a slow increase of the characteristic length is a typical feature of glasses.
Although the origin of the very slow dynamics in our model is basically the same as in the SS
model, the nature of the glassy transition is quite different.
Firstly, above the glassy transition the model does not enter the fast-dynamics regime (as the SS
model does) but remains trapped in the supercooled liquid phase. 
In addition, our estimation of a certain characteristic time shows that energy barriers exist even
above the glassy transition.
It strongly suggests that the glassy transition in our model is not induced by a corner-rounding
transition and that energy barriers persist even above the glassy transition.

There are other properties of the model which are very interesting.
Our simulations show that in this model in a certain temperature range, depending on the initial
conditions, the model might be either in the liquid or the crystal phase.
Using the thermodynamic integration method we calculated the free energies of both phases and, as
expected, the crossing point (i.e., the first-order phase transition) appears approximately in the 
middle of this temperature range.
Such metastable effects (hysteresis) frequently accompany first-order phase 
transitions~\cyt{PENROSE,RIKVOLD}.
A distinctive feature of metastability of our model is that this is an extremely strong effect.
We show that even for temperatures close to the limits of hysteresis only a large droplet of
the stable phase injected into the metastable phase can divert the evolution of the system into the stable phase.
Spontaneous nucleation of such large droplets is an extremely improbable event, and is well beyond
computational capacities of modern computers.

In addition, our simulations show that upon cooling the glassy transition is accompanied by a
pronounced peak in the specific heat.
This (pseudo-)critical behaviour gives rise to certain slow modes in the liquid phase of our model.
Such slow modes for glassy systems are predicted by the Mode-Coupling Theory~\cyt{GOTZE} and are
experimentally verified as so-called $\alpha$-oscillations~\cyt{ALPHA}.

Our paper is organised as follows.
In section 2 we introduce the model, review its basic properties and study the evolution of random 
quench.
In section 3 we calculate the free energy and the specific heat and present the time evolution of
internal energy.
In section 4 we discuss metastability properties of our model.
Section 5 contains our conclusions.
\SECTION{}{Basic properties and domain coarsening}
The model is defined by the following Hamiltonian
\beq
H = -\sum S_iS_jS_kS_l,
\eeq {1}
where $S_i=\pm 1$ and summation is performed over all elementary plaquettes ($i,j,k,l$) of a
simple cubic lattice of linear size $L$ with spins placed at the {\it sites} of the lattice.
Model~\eq{1} is a special case of the so-called gonihedric models, which have recently
been studied in the context of the lattice field theory~\cyt{SAV,BAIG}.
Closely related models have been studied in the context of random
surface models and a very rich phase structure mapped out~\cyt{KAROWSKI,MARITAN}.

The basic properties of model~\eq{1} which are already known can be briefly described as follows:
The ground state is strongly degenerate with the degeneracy $\sim 2^{3L}$ (the ground state entropy
per site is thus zero).
Monte Carlo simulations show~\cyt{BAIG,LIPO} that upon heating of an arbitrary ground state
(crystal) configuration, the model undergoes a discontinuous phase transition into the disordered
phase (liquid) around $T\sim 3.9$; the temperature scale is set as in~\cyt{LIPO} with
Boltzmann constant put to unity.
An important feature of model~\eq{1} is the shape dependence of excitations~\cyt{LIPO}: it is not
only the size of the excitation which determines its energy (as in the ordinary two-spin Ising
model) but also its shape (see~\fig{cubes}). 
This property, which holds also for the SS model, gives rise to energy barriers which are in turn
responsible for the very slow dynamics of both models.

To study the evolution of the random configuration quenched to low temperature, we measured the
energy excess $\delta E(t)=E(t)-E_0$ over the ground state energy $E_0=-3$.
One expects~\cyt{SHORE} that the inverse of this quantity sets the
characteristic length scale $l(t)$ of the system, which roughly corresponds to the average
size of domains.
Moreover, there is convincing evidence~\cyt{BRAY} that in many systems with nonconservative
dynamics and a scalar order parameter (i.e., conditions which are satisfied in our approach)
$l(t)$ increases asymptotically in time as $l(t)\sim t^n$ and $n=1/2$.
However, in some systems $l(t)$ is known to increase much more slowly in time,
namely logarithmically $l(t)\sim {\rm log}(t)$.
These exceptional systems include some random (at the level of the
Hamiltonian) systems~\cyt{LAI,RAO}, and the SS model for temperatures below the
corner-rounding transition~\cyt{RAO1}.
It is the energy barriers developing in these systems during the coarsening which
cause such a slow increase of $l(t)$.

We performed standard~\cyt{BINDER} Monte Carlo simulations using the Metropolis algorithm with a random
sequential update.
Unless stated otherwise, periodic boundary are imposed.
The log-log plot of $1/\delta E(t)$ as a function of time for model~\eq{1} is shown in~\fig{f1}.
The presented results are obtained for $L=40$ but very similar behaviour was observed for
$L=30$.
From~\fig{f1} it is clear that for $T=1.5$ and 2.8 the asymptotic slopes of the curves are much
smaller than 1/2 and there is a tendency for these curves to bend downwards.
Taking into account the absence of models with $n$ considerably smaller than 1/2 and the existence
of energy barriers in model~\eq{1} of basically the same nature as in the SS model, suggests that
for the examined temperatures the characteristic length asymptotically increases logarithmically in
time.
We cannot exclude, however, that in this temperature regime the increase of $l(t)$ is even more
exotic, with neither logarithmic nor power-law increase.
It would appear that  such a slow increase of $l(t)$ takes place even for $T=3.3$ and 3.4,
but the behaviour of $l(t)$ for these temperatures is obscured by the metastability
effects, since before collapsing into the glassy phase the system 
remains in the liquid state for some time.

The difference between our model and the SS model becomes clear when 
we approach the glassy transition (which we roughly estimate to take place at
$T=T_{{\rm g}} \sim 3.4$)
by {\it increasing} the temperature. 
In the SS model for temperatures below the critical point but above the corner-rounding transition 
thermal fluctuations roughen corners of domains and energy barriers are no longer relevant.
Consequently, the ``ordinary '' dynamics with $n=1/2$ is restored and the system rapidly evolves
toward the low-temperature phase.
On the contrary, in model~\eq{1} for $3.4<T<3.9$ the random quench does not even
evolve toward the low-temperature phase but remains disordered~\cyt{LIPO}.
Since the low-temperature phase does exist at these temperatures (as we have already mentioned, the
transition from the low-temperature phase to the disordered phase takes place around $T\sim 3.9$),
there should be some barriers which prevent the liquid from collapsing.
In our opinion, these barriers are of entropic origin and they are probably related
to the strong degeneracy of the ground state, which would explain why the behaviour of our
model is different from the SS model~\cyt{ENTROPY}.
This phenomenon is discussed further in section 4.

\SECTION{freeenergy}{Free energy and specific heat}
The free energy encodes all the important thermodynamic information about a system.
In equilibrium statistical mechanics this quantity is defined as 
\beq
f=-T/N\ln [\sum {\rm e}^{-H/T}],
\eeq{free}
where $N$ is the number of particles (lattice sites) and $H$ is the Hamiltonian of the system with 
the Boltzmann constant put to unity.
However, calculation of the above defined free energy using Monte Carlo simulations is not
straightforward and requires thermodynamic integration~\cyt{BINDER}.
To calculate the free energy of our model in the liquid and crystal phases we used the following
equations:
\beq
f_{{\rm cryst}}=u-T\int_{0}^{T} \frac{c}{T} dT, \ \ \ 
f_{{\rm liq}}= -Ts(\infty) + T\int_{0}^{1/T} u d(\frac{1}{T}).
\eeq{2}

In the above equation $c$ and $u$ denote the specific heat and internal energy, respectively 
(calculated using standard formulae~\cyt{BINDER}), and $s(\infty)=\ln (2)$ is the entropy per site
at infinite temperature.
The results of our calculations are shown in~\fig{f5}.
We checked the stability of our results with respect to the integration step $\Delta T$ (or
$\Delta (1/T)$), the system size $L$ and the number of Monte Carlo steps at each temperature.

In~\fig{f5} one can see that the free energies of the crystal and of the liquid cross around
$T=3.6$ and we expect that this is the temperature of the first-order phase transition.
The estimation of the transition temperature is in a good agreement with calculations done using 
the Cluster Variational Method~\cyt{CIRILLO}.
However, due to the strong metastability, which is discussed in more detail in the next section, 
the transition at this point is very difficult to observe.
Indeed, if we heat a crystal sample, the transition occurs around $T=3.9$, while cooling a
liquid sample results in the glassy transition around $T=3.4$.

To overcome the metastability and confirm that a first-order transition does take place around
$T=3.6$, we simulated the system with a nonuniform initial configuration.
Namely, we prepared the system with one half of it in the crystal phase (e.g., all spins
'up') and the second half in the liquid phase (all spins random); see~\fig{initial}.
Such a choice facilitates the evolution toward the stable phase (i.e., the one with the lowest free
energy), because both phases are present in the initial configuration and the system does not have
to nucleate the stable phase.
Results of such simulations are shown in~\fig{halfhalf}.
One can clearly see that the evolution of the system depends on whether the temperature is above
or below $T=3.6$.
For $T< (>)3.6$ the crystal (liquid) phase gradually expands until the stable phase invades the
whole system.
The identification of the final state is obtained from a comparison of its internal energy with
simulations which use a uniform initial state and also from the visual inspection of snapshot
Monte Carlo configurations.

To provide additional information about model~\eq{1}, we measured the variance of the internal
energy and calculated the specific heat~\cyt{BINDER}.
Our data is shown in~\fig{heat}.
When we start our simulations from the ground-state configuration (heating), the behaviour of
the specific heat confirms a transition around $T=3.9$, in agreement with earlier 
simulations~\cyt{BAIG,LIPO}.
However, under cooling this peak is shifted toward much lower temperature and it seems to coincide
with our estimation of $T_{{\rm g}}$.
Under cooling we do not observe any singularity in the specific heat until $T=T_{{\rm g}}$, which
indicates that during cooling and for $T>T_{{\rm g}}$ the system remains in the liquid phase.
Let us notice that the locations of both peaks are almost unchanged after increasing the system
size by almost a factor of two.
Moreover, for $T<T_{{\rm g}}$ the specific heat is slightly larger upon cooling than upon
heating.
This is in agreement with the fact that in this temperature range the model has slow dynamics and
cannot reach (within numerically accessible computing time) the crystal phase, for which the 
specific heat is very small.

Let us note that the pronounced peak in the specific heat presumably indicates critical
or pseudo-critical behaviour in model~\eq{1}.
If so, we might expect that certain relaxation times might substantially increase or even diverge,
which should be observed as some slow modes in the system.
In~\fig{f3} we present the time dependence of the energy $E(t)$.
Although an explicit calculation of the Fourier transform of this quantity would be desirable, it
is rather clear that for $T=3.42$, which is very close to $T_{{\rm g}}$, in addition to fast
fluctuations the system exhibits slow fluctuations with a time scale $\sim 150$.
For higher temperatures ($T=3.6$) the time scale of slow fluctuations decreases and eventually
($T=3.8$) becomes hardly distinguishable from fast oscillations.
Such slow and fast modes resemble experimentally observed $\alpha-$ (slow) and
$\beta-$ (fast) oscillations in real glasses~\cyt{ALPHA}.
It would be interesting to check to what extent the properties of our model agree with the Mode
Coupling Theory~\cyt{GOTZE}, according to which such slow modes are a key factor driving a
glassy transition.
\SECTION{Metas}{Metastability}
Results presented in the previous sections suggest that  in the temperature range $T_{{\rm
g}}<T<T_{{\rm c}}$ the system might remain either in the liquid or the crystal phase.
In the present section we present some additional results concerning the (meta-)stability of 
the liquid and crystal phases.
\SUBSECTION{tau}{Characteristic times}
We measured various characteristic times imposing different initial and boundary conditions and
monitoring the evolution toward a final state.
To check the stability of the liquid ($\tau_{{\rm liq}}$), we used a random initial configuration
and simulated the system until the energy reached $E=-2.3$.
\footnote{This value is chosen rather
arbitrarily, but once the system reaches this energy it does not return to the
liquid phase.}
To calculate $\tau_{{\rm liq}}$ we made 100 independent runs.
Our results for $T=3.5$ are shown in~\fig{f2} and they suggest that the escape time
increases at least exponentially with the system size.

To check the stability of the crystal, one should measure the time
($\tau_+$) needed for the crystal to be transformed into the liquid.
It would be particularly interesting to examine the size dependence of $\tau_+$ for $3.6<T<3.9$,
i.e., for temperatures where the crystal is metastable.
We have found, however, that this quantity increases very rapidly with the system size and in
this temperature range it is virtually impossible to increase the system size beyond $L=6$.
The stability of this phase might be also inferred from other measurements we made.  
in which we estimated the time ($\tau_{ +-}$) needed to shrink a cubic excitation of
size $L$.
This technique parallels that which has already been applied to the SS model~\cyt{SHORE}: the
initial configuration has ``up ''spins at the boundary of the cube of size $L+2$ (which are kept
fixed) and ``down ''spins inside this cube.
Simulations are performed until the magnetization of the interior of the cube decays to zero and the
time needed for such a run is recorded.
To calculate $\tau_{+-}$ at a given temperature we made 100 independent runs.
Our results for $T=3.6$ (\fig{f2}) suggest that $\tau_{+-}$ increases approximately exponentially
with $L$.
Such an increase confirms that the glassy transition in model~\eq{1} is not induced by the
corner-rounding transition since 
above the corner-rounding transition one expects $\tau_{+-}\sim
L^2$~\cyt{SHORE} and the data in~\fig{f2} should bend considerably downwards.
It also confirms the stability of the crystal since it is clear that bringing the crystal
(a homogeneous, low-energy  configuration) into the liquid is a slower process than shrinking an
excitation.
Similar size dependence of $\tau_{{\rm liq}}$ and $\tau_{+-}$ was also observed for other
temperatures in the interval $3.4<T<3.9$.

We also measured the characteristic times $\tau_{{\rm liq}}$ and $\tau_+$ outside the interval
$3.4<T<3.9$.
Simulations were done for several values of system size $L$ and the results were extrapolated to
the thermodynamic limit ($L\rightarrow \infty$) using a simple fit~\cyt{COMM2}.
These extrapolated values are shown in~\fig{fdod}.
One can see that outside this temperature range $\tau_{{\rm liq}}$ and $\tau_+$ are definitely 
finite and they seem to diverge upon approaching $T=3.4$ from below ($\tau_{{\rm liq}}$) and
$T=3.9$ from above ($\tau_+$).

\SUBSECTION{droplets}{Droplets of a stable phase}
The numerical data  presented in the previous subsection suggests that for $3.4<T<3.9$ the model has two
different phases of effectively infinite life-time.
Such a result would be in disagreement with the result that in short-range interacting
systems metastability is only a quantitative effect~\cyt{PENROSE,RIKVOLD}.
In this section we show, however, that data presented in~\fig{f2} and~\fig{fdod} are misleading and in the
thermodynamic limit ($L\rightarrow \infty$) the characteristic times $\tau_{{\rm liq}}$ for
$T<3.6$ and $\tau_+$ for $T>3.6$ should be finite.
The time scale of these metastable effects is, however, enormously long in comparison
with the length of our simulations. 

One expects that metastable phases have only a finite life-time due to droplet nucleation.
When a sufficiently large droplet of stable phase nucleates inside a metastable phase, it
diverts evolution of the system toward a stable phase.
Since the critical (i.e., minimal) droplet size is finite, there is a finite probability of
spontaneous nucleation of such droplets and thus a life-time of a metastable phase is also finite.

To check whether such a mechanism operates in model~\eq{1}, we monitored the evolution of the
system
with droplets introduced by hand into the initial configuration (see~\fig{initial}).
Our results for $T=3.5$ are shown in~\fig{droplet}.
One can see that when a droplet of the crystal is sufficiently small (of linear size $M=18, 24$),
the system after some transient ends up in the liquid phase.  
However, a large droplet of size $M=36$ diverts evolution of the system
toward the more stable (crystal or glassy) phase.
We performed similar simulations to examine the (meta-)stability of the crystal phase.
Setting $T=3.8$ we observed that droplets of liquid of size $M\geq 24$ divert evolution of
the crystal phase toward the stable liquid phase.

The above results show that the droplet-nucleation mechanism is effective in model~\eq{1} and
metastable phases are of finite life-time.
However, the important question is how long is this life-time.
This quantity is determined by the inverse of the probability of the spontaneous nucleation of
critical droplets.
It is clear that for computationally accessible systems ($L\sim 100$) spontaneous
nucleation of crystal droplets of linear size $M\sim 30$ is an extremely unlikely event,
which takes place on astronomical time scales.
We should emphasize that it does not mean that our model predicts such a life-time of metastable liquids.
Since the nucleation of droplets is basically a local event, its probability for macroscopic
systems increases merely due the system size (droplets might nucleate independently in many
places).

The radius of critical droplets presumably vanishes upon approaching the limits of hysteresis 
(i.e., $T=3.4$ and 3.9).
Thus, very close to these limits one should be able to observe finite-life-time effects 
such as
the spontaneous collapse of liquid into the crystal (or maybe glassy) phase.
\SECTION{conclusions}{Conclusions}
In the present paper we have 
studied the three-dimensional Ising model with four-spin interactions.
The Hamiltonian of this model is homogeneous, non-frustrated and contains only short-range
(plaquette) interactions.
Nevertheless, we found that this model has very interesting dynamical and thermodynamical
properties.
In particular, our numerical results suggest that the model has a very slow coarsening dynamics in its
low-temperature phase.
Moreover, due to very strong metastability, in a certain temperature range the model can
remain either in a crystal or liquid phase depending on how it has been prepared.
We have shown that a droplet-nucleation mechanism is effective in this model and thus that
metastability in this model is a finite-size (but very strong) effect.
The time scale of spontaneous nucleation for critical droplets is extremely large and
well beyond the timescales of our simulations.
Upon cooling the metastable liquid collapses into a glassy phase.
We have found that the glassy transition in our model, which is purely dynamical in nature, is preceded by a 
pronounced peak in the specific heat (pseudo-critical behaviour) and also by slow oscillations of the 
internal energy.
\begin {thebibliography} {00}

\bib {GOTZE} W.~G\"{o}tze, in {\it Liquid, Freezing and Glass Transition}, Les Houches Summer
School, ed. J.~P.~Hansen, D.~Levesque and J.~ Zinn-Justin (North-Holland, Amsterdam, 1989).
C.~A.~Angell, Science {\bf 267}, 1924 (1995).

\bib {JPHYSC} For a collection of recent papers on physics of glasses see the special issue of
J.~Phys.~Condens.~Matt.~{\bf 11} no 10A (1999).

\bib{KOB} Recently, however, important progress in studying off-lattice models has been made; see,
e.g., W.~Kob, in {\it Annual Reviews of Computational Physics}, ed.~D.~Stauffer (World Scientific,
Singapore, 1995) vol.~III.
W.~Kob and H.~C.~Andersen, Phys.~Rev.~E {\bf 52}, 4134 (1995).

\bib {MEANFIELD} S.~Franz and J.~Hertz, Phys.~Rev.~Lett.~{\bf 74}, 2114  (1994).
J.~P.~Bouchaud and M.~Mezard, J.~Phys.~I (France) {\bf 4}, 1109 (1994).

\bib {SS} J.~D.~Shore and J.~P.~Sethna, Phys.~Rev.~{\bf B 43}, 3782 (1991).

\bib {SHORE} J.~D.~Shore, M.~Holzer and J.~P.~Sethna, Phys.~Rev.~{\bf B 46}, 11376 (1992).

\bib {LIPO} A.~Lipowski, J.~Phys.~{\bf A 30}, 7365 (1997).

\bib {PENROSE}  O.~Penrose and J.~L.~Lebowitz, in {\it Fluctuation Phenomena}, ed. E.~W.~Montroll
and J.~L.~Lebowitz (Amsterdam: North Holland, 1979).

\bib{RIKVOLD} P.~A.~Rikvold and B.~M.~Gorman, in {\it Annual Review of Computational Physics}
vol.I, ed.~D.~Stauffer (Singapore: World Scientific, 1994).

\bib {ALPHA} F.~H.~Stillinger, Science~{\bf 267}, 1935 (1995).
 
\bib {SAV} R.~V.~Ambartzumian, G.~S.~Sukiasian, G.~K.~Savvidy and K.~G.~Savvidy, Phys.Lett.~{\bf B
275}, 99 (1992).

\bib {BAIG} D.~Espriu, M.~Baig, D.~A.~Johnston and R.~P.~K.~C.~Malmini, J.~Phys.~{\bf A 30}, 405
(1997).

\bib {KAROWSKI} M.~Karowski, J.~Phys.~A {\bf 19}, 3375 (1986).

\bib {MARITAN} A.~Cappi, P.~Colangelo, G.~Gonella and A.~Maritan, Nucl.~Phys.~B {\bf 370}, 659
(1992);\\ 
P. Colangelo, G. Gonnella and A. Maritan, Phys. Rev. {\bf E47}, 411 (1993);\\     
G. Gonnella and A. Maritan, Phys. Rev. {\bf B48}, 932 (1993);\\
G.Gonnella, S.Lise and  A. Maritan, Europhys. Lett. {\bf 32}, 735 (1995).           

\bib {BRAY} A.~J.~Bray, Adv.~in Phys.~{\bf 43}, 357 (1994).

\bib {LAI} Z.~W.~Lai, G.~F.~Mazenko and O.~T.~Valls, Phys.~Rev.~B {\bf 37}, 9481 (1988).

\bib{RAO} M.~Rao and A.~Chakrabarti, Phys.~Rev.~Lett.~{\bf 71}, 3501 (1993).

\bib{RAO1} For additional confirmation of logarithmically slow growth of $l(t)$ in the SS model see
also: M.~Rao and A.~Chakrabarti, Phys.~Rev.~E {\bf 52}, R13 (1995).

\bib {BINDER} K.~Binder, in {\it Applications of the Monte Carlo Method in Statistical Physics},
ed.~K.~Binder, (Berlin: Springer, 1984).

\bib {ENTROPY} Entropic barriers are known to play an important role in glassy systems: see, e.g.,
U.~Mohanty, I.~Oppenheim and C.~H.~Taubes, Science~{\bf 266}, 425 (1994).
T.~R.~Kirkpatrick and P.~G.~Wolynes, Phys.~Rev.~{\bf B 36}, 8552 (1987).

\bib {CIRILLO} E.~N.~M.~Cirillo, G.~Gonella, D.~A.~Johnston and A.~Pelizzola, Phys.~Lett.~{\bf A
226}, 59 (1997).

\bib {COMM2} We plotted the results for finite $L$ as a function of $x=1/L$ and then used parabolic
fit to extrapolate the value at $x=0$.

\end {thebibliography}
\begin {figure_captions}

\elem {cubes} In model~\eq{1} the energy of an excitation (e.g., 'down' spins
surrounded by 'up' spins) is proportional to the total length of edges of the boundary of that
excitation.
To remove a cubic excitation (a), the system is likely to proceed through configurations like that
shown in (b) and (c).
It is easy to realize that in (b) and (c) the total length of edges and thus the energy of such 
configurations is larger than that in (a).
In the case (c) the energy increase is proportional to the linear size of the excitation.
On the contrary, in a two-spin Ising model all configurations would have the same energy (since
the area of all configurations is the same) and removing of excitations would proceed without
climbing any energy barriers.

\elem {f1} The log-log plot of of $1/\delta E(t)$ as a function of time $t$ ($L=40$).
The dashed line has a slope 1/2.

\elem {f5} The free energy of liquid ({\large +}) and crystal ($\diamond$).
Calculations were done for $L=30$, and 5000 Monte Carlo steps were used at each temperature.
The integration steps were $\Delta T = \Delta (1/T) = 0.025$.

\elem {initial} Two-dimensional sections of initial configurations used in our calculations.
(a) An initial configuration used in the calculations of internal energy shown in~\fig{halfhalf}.
(b) An initial configuration with a droplet of crystal phase injected into the liquid
phase (see Section 4.2).
Spins in the crystal/liquid part of the system are initially set 'up'/at random.

\elem {halfhalf} The internal energy $U$ as a function of time for $L=50$ and an initial
configuration composed of 50\% of crystal and 50\% of liquid (see~\fig{initial}a).
The steady-state values of $U$ for $T=3.5,3.55$ and for $T=3.65,3.7$ are identical (within error
bars) with internal energy of crystal and liquid at corresponding temperatures. 

\elem {heat} The specific heat calculated during (i) heating for $L=24$ ($\diamond$) and $L=40$
($\times$) (ii) cooling for $L=24$ ({\large +}) and $L=40$ ($\Box$).
For each temperature and system size we made runs of $10^4$ Monte Carlo steps plus $10^3$ Monte
Carlo steps for relaxation.

\elem {f3} The time dependence of the energy $E(t)$ in the liquid phase at various
temperatures ($L=40$).

\elem {f2} The size dependence of the logarithm of the escape times $\tau_{{\rm liq}}$
($\Box$) and $\tau_{+-}$ ({\large +}).
Calculation of $\tau_{{\rm liq}}$ and $\tau_{+-}$ was done for $T=3.5$ and $T=3.6$, respectively.

\elem {fdod} The inverse of the characteristic times $\tau_{{\rm liq}}$ ({\large +}) and $\tau_+$
($\diamond$) as a function of temperature.
The plotted results are obtained by extrapolation of the finite-size data to the thermodynamic
limit.

\elem {droplet} The internal energy $U$ as a function of time for $T=3.5$.
The initial configuration consists of a droplet of crystal phase of size $M$ injected into the
liquid phase.

\end {figure_captions}
\end {document}